# On proportional fairness of uplink spectral efficiency in cell-free massive MIMO systems

Viet Quoc Pham[1,2] | Ha Hoang Kha[1,2] | Le Ty Khanh[1,2]

[1]Faculty of Electrical and Electronics Engineering, Ho Chi Minh City University of Technology (HCMUT), Ho Chi Minh City, Vietnam

[2]Vietnam National University Ho Chi Minh City, Ho Chi Minh City, Vietnam

**Summary**

This paper is concerned with the proportional fairness (PF) of the spectral efficiency (SE) maximization of uplinks in a cell-free (CF) massive multiple-input multiple-output (MIMO) system in which a large number of single-antenna access points (APs) connected to a central processing unit (CPU) serve many single-antenna users. To detect the user signals, the APs use matched filters based on the local channel state information while the CPU deploys receiver filters based on knowledge of channel statistics. We devise the maximization problem of the SE PF, which maximizes the sum of the logarithm of the achievable user rates, as a jointly nonconvex optimization problem of receiver filter coefficients and user power allocation subject to user power constraints. To handle the challenges associated with the nonconvexity of the formulated design problem, we develop an iterative algorithm by alternatively finding optimal filter coefficients at the CPU and transmit powers at the users. While the filter coefficient design is formulated as a generalized eigenvalue problem, the power allocation problem is addressed by a gradient projection (GP) approach. Simulation results show that the SE PF maximization not only offers approximately the achievable sum rates as compared to the sum-rate maximization but also provides an improved trade-off between the user rate fairness and the achievable sum rate.

**KEYWORDS**

cell-free massive MIMO, gradient projection, proportional fairness, spectral efficiency

## 1 | INTRODUCTION

Cell-free (CF) massive multiple-input multiple-output (MIMO) networks have been considering as a promising technology for next-generation wireless systems thanks to their potential improvement of the spectral efficiency (SE) and energy efficiency (EE).[1] In these networks, a large number of distributed access points (APs) equipped with single or multiple antennas coherently serve a much smaller number of users at the same time and in the same frequency band. As an integration of distributed MIMO and massive MIMO techniques, CF massive MIMO networks can adhere all benefits of two techniques.[2] Marzetta[3] showed that the deployment of linear processing in massive MIMO networks is nearly optimal thanks to the favorable propagation and channel hardening, which results in lowering the complexity of signal processing at the mobile stations (MSs). Also, by distributing the APs over a large area, antennas are brought close to the users, which leads to a high degree of macrodiversity and low path losses.[2] Therefore, CF massive MIMO networks could provide widely good quality-of-service (QoS).

The studies on CF massive MIMO networks have recently been an active research topic. Ngo et al[2] introduced power control algorithms to maximize the minimum user rate while guaranteeing uniformly good service at all users. It demonstrated that the CF massive MIMO network outperforms the conventional small-cell system in terms of per-user throughput.[2] In another work, Ngo et al[1] analyzed the effects of the backhaul power consumption and maximized the total EE of the downlinks (DLs) in CF massive MIMO networks under the constraints of per-user SE and transmit power at each AP. Bashar et al[4] investigated the max-min signal-to-interference-plus-noise ratio (SINR) problem of uplinks (ULs) in CF massive MIMO systems. The authors introduced receiver filter coefficients at a central processing unit (CPU) and proposed geometric programming to obtain the optimal power allocation. Simulation results[4] indicated that adding the receiver filter coefficients substantially improved the UL rates of the network. Also, Nguyen et al[5] considered the problem of load balancing and power control in the ULs of a CF massive MIMO system. The authors[5] introduced a low-complexity method for solving the load balancing problem and proposed a successive approach to maximize the sum SE of the network. Simulation results[5] revealed that, in the high QoS regime, the proposed approach can provide better SE than the maximum signal-to-noise ratio (max-SNR) association in which users are only served by the highest SNR base station (BS). Vu et al[6] studied a CF massive MIMO network in which a very large number of full-duplex (FD) multiple-antenna APs simultaneously serve many half-duplex single-antenna users. The numerical results[6] confirmed that the FD CF massive MIMO systems offer higher SE than the FD collocated massive MIMO systems. Nguyen et al[7] proposed a low-complexity power control algorithm maximizing the DL EE of a CF massive MIMO network in which the backhaul power consumption and the imperfect channel state information (CSI) have been taken into account. Interdonato et al[8] proposed a closed-form expression of the DL SE of a CF massive MIMO system with multiantennas APs using the full-pilot zero-forcing (fpZF) precoding scheme.

To achieve a good balance between maximizing the sum rate and maintaining fairness among users in a network, a proportional fairness (PF) utility function should be considered.[9-11] Indeed, the fairness problem among the users may arise when maximizing the sum rate of a network due to different distances from the users to the BS.[10] The PF utility function was introduced by Kelly[12] and later applied in various studies; see, for example, by Lau,[9] Cirik,[10] and Hadzi-Velkov et al[11] and references therein. Lau[9] generalized Qualcomm's original PF scheduler to a generic wireless communication network with a multiantenna BS. Simulation results[9] confirmed that the proposed PF scheduler provided a good balance between the fairness and the sum throughput of the system. Cirik[10] considered the fairness problem of a cellular network in which a FD BS simultaneously serves several UL and DL users. A gradient projection (GP) method was proposed to solve the problem of maximizing the sum of the logarithm of the UL and DL rates with power constraints at the BS and the UL users. It was demonstrated that the proposed approach[10] achieved a good balance between maximizing the sum rate and enhancing fairness among users. Hadzi-Velkov et al[11] considered wireless powered communication networks that employ the slotted ALOHA protocol. By exploiting the random access (RA) nature of slotted ALOHA, the authors proposed a proportionally fair resource allocation scheme that optimizes transmit power at the BS and duration of the energy harvesting (EH) and the RA phases in terms of ensuring fairness among users.

In this paper, we study the maximization problem of the SE PF of the uplinks in a CF massive MIMO network, which aims at maximizing the sum rate and mitigating the achievable user rate gaps among the users. It should be noted that Nguyen et al[5] maximized sum SE of a CF massive MIMO system. Ngo et al[2] considered the max-min SINR problem in a CF massive MIMO system to maximize the smallest SINR. The similar problem was also considered by Bashar et al[4]; however, the authors introduced receiver filter coefficients at the CPU to improve the UL rates of the networks and proposed a new approach solving the max-min SINR problem. All of the aforementioned works focused on the sum-rate maximization (SRM) or the max-min SINR problems. To best of our knowledge, there has been no previous work on the SE PF in CF massive MIMO systems. Indeed, the notion of PF has been proposed and taken into account previously[9-11] for different scenarios rather than in CF massive MIMO networks. More specifically, the PF utility function was studied by Cirik[10] for designing the precoders in FD MIMO systems which are different from our present work which designs the receiver filters and transmit power allocation. Likewise, Lau[9] adopted the PF utility function for space-time scheduling in wireless communications. However, the objective function in the work[9] is concave, and the design problem therein is a convex optimization problem. Therefore, being motivated from previous works, in this paper, an iterative algorithm is proposed to solve the nonconvex optimization problem which designs the receiver filter coefficients at the CPU and power allocation at the user to maximize the SE PF utility function. The formulated design problem can be solved by alternatively finding optimal filter coefficients at the CPU and transmit powers at the users. While the filter coefficient design problem is a generalized eigenvalue problem, the power allocation problem is resolved by a GP approach.[13,14] We derive an iterative algorithm based on GP to maximize the sum of the logarithms of the UL achievable rates, subject to transmit power constraints at the users. Simulation results are provided to illustrate the performance of the proposed algorithm. Our major contributions are summarized as follows:

- We formulate the filter coefficient design and power allocation in the ULs of CF massive MIMO systems as the SE PF maximization (SEPFM) subject to transmit power constraints. Then, we propose an iterative algorithm in which optimal solutions can be achieved by alternatively solving generalized eigenvalue and power allocation problems. To tackle with the challenges of solving the power allocation problem, we develop an iterative GP method with the guarantee of convergence.
- By numerical results, we verified the convergence of the proposed algorithm. In addition, we numerically analyze the performance of the network in three different optimization approaches, namely, the SEPFM, the SRM, and the max-min SINR.[4] It is shown that by incorporating the PF notion into the SE maximization problem, CF massive MIMO networks can not only achieve a remarkable sum rate but also guarantee a relatively fair distribution of user rates.

*Outline:* The remainder of this paper is organized as follows. In Section 2, we describe a model of the CF massive MIMO system with signal processing at the APs and CPU. In Section 2, we present the achievable UL rates and the SEPFM problem. The receiver filter coefficient design and power allocation are devised in Section 3. Simulation results and discussion are given in Section 4. Finally, we conclude the paper in Section 5.

*Notation:* Boldface letters denote column vectors. The superscripts $()^*, ()^T$, and $()^H$ stand for the conjugate, transpose, and conjugate-transpose, respectively. The Euclidean norm and the expectation operators are denoted by $\|.\|$ and $\mathbf{E}[.]$, respectively. $\odot$ and $\oslash$ represent the Hadamard product and division, respectively. Finally, $c \sim \mathcal{CN}(0, \sigma^2)$ denotes a circularly symmetric complex Gaussian random variable (RV) $c$ with zero mean and variance $\sigma^2$, and $r \sim \mathcal{N}(0, \sigma^2)$ denotes real valued Gaussian RV $r$.

## 2 | SYSTEM MODEL

Consider the UL transmission in a CF massive MIMO network[2] in which $M$ single-antenna APs connected to a CPU serve $K$ single-antenna users as illustrated in Figure 1. The set of APs is denoted by $\mathcal{M} = \{1, 2, ..., M\}$ while the set of all users is denoted by $\mathcal{K} = \{1, 2, ..., K\}$. The APs and the users are randomly distributed in a wide coverage area. In general, a CF massive MIMO system operates in the time duplex division (TDD) with three fundamental phases: UL training phase, DL data transmission, and UL data transmission. However, in this paper, we focus on the UL process in which APs estimate CSI based on received pilot sequences transmitted from the users in the UL training phase, and then, data will be transmitted from the users to the APs. Denote the channel coefficient between user $k$ and AP $m$ by $h_k^m$ which is modeled as[2]

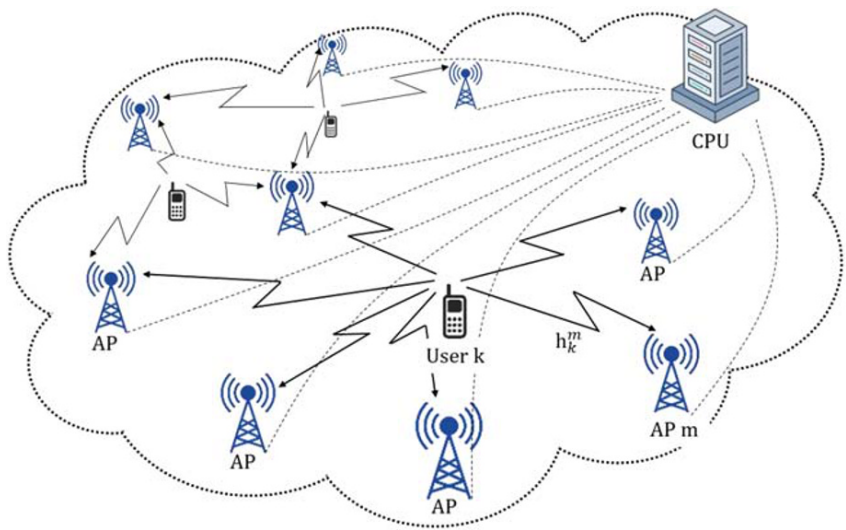

**FIGURE 1** The uplink transmission model of a CF massive MIMO network

$$h_k^m = \left(\beta_k^m\right)^{1/2} \tilde{h}_k^m, m \in \mathcal{M}, k \in \mathcal{K}, \tag{1}$$

where $\tilde{h}_k^m$ represents the small-scale fading which is assumed to be independent and identically distributed (i.i.d) RV with zero mean and unit variance, $\mathcal{CN}(0,1)$ while $\beta_k^m$ denotes the large-scale fading.

*UL training phase*: Let the length of the coherence interval (in samples) of the UL channels be $\tau_c$. To estimate CSI at the APs, the pilot sequences in the duration of $\tau$ symbols ($\tau < \tau_c$) are simultaneously transmitted from the users to the APs for every coherence interval. By following the previous approaches,[2,4,15] pilot sequences from a given set are randomly assigned to the users. Let the pilot sequence transmitted by user $k$, $k \in \mathcal{K}$, be $\sqrt{\tau}\varphi_k \in \mathbb{C}^{\tau \times 1}$, where $\|\varphi_k\|^2 = 1$. Then, the $\tau \times 1$ received pilot vector at AP $m$ can be formulated as

$$\mathbf{r}_p^m = \sqrt{\tau\rho_p} \sum_{k \in \mathcal{K}} h_k^m \varphi_k + \mathbf{n}_p^m, \tag{2}$$

where $\rho_p$ and $\mathbf{n}_p^m$ denote the normalized signal-to-noise ratio (SNR) of each pilot symbol and a vector of additive noise at AP $m$, respectively. The elements of $\mathbf{n}_p^m$ are assumed to be i.i.d RVs with zero means and unit variances, $\mathcal{CN}(0,1)$. Then, AP $m$ uses the pilot sequence $\varphi_k$ to decorrelate the received signal, $\mathbf{r}_p^m$, as follows[2]:

$$\check{r}_{pk}^m = \varphi_k^H \mathbf{r}_p^m = \sqrt{\tau\rho_p} h_k^m + \sqrt{\tau\rho_p} \sum_{k' \in \mathcal{K} \setminus \{k\}} h_{k'}^m \varphi_k^H \varphi_{k'} + \varphi_k^H \mathbf{n}_p^m. \tag{3}$$

Then, AP $m$ uses a minimum mean-square error (MMSE) estimator to estimate channel $h_k^m$. Accordingly, the linear MMSE estimate of $h_k^m$ given $\check{r}_p^m$ can be obtained by[2,16]

$$\hat{h}_k^m = \frac{\mathbf{E}\left\{\left(\check{r}_{pk}^m\right)^* h_k^m\right\}}{\mathbf{E}\left\{\left|\check{r}_{pk}^m\right|^2\right\}} \check{r}_{pk}^m = \vartheta_k^m \check{r}_{pk}^m, \tag{4}$$

where

$$\vartheta_k^m \triangleq \frac{\sqrt{\tau\rho_p} \beta_k^m}{\tau\rho_p \sum_{k' \in \mathcal{K}} \beta_{k'}^m \left|\varphi_k^H \varphi_{k'}\right|^2 + 1}. \tag{5}$$

Here, we assume that the APs to have knowledge of large-scale fading of the users,[2] that is, $\beta_k^m, m \in \mathcal{M}, k \in \mathcal{K}$. The channel estimates of small-scale fading in (4) are locally used to design receiver filter coefficients at the CPU and allocate transmit powers at users towards maximizing the SE PF of the network.

*UL data transmission:* In the UL data transmission, all $K$ users are supposed to simultaneously send their data to the APs. The transmitted signal from user $k$ is

$$s_k = \sqrt{p_k} x_k, \tag{6}$$

where $x_k, \mathbf{E}\left[|x_k|^2\right] = 1$ be a data symbol of user $k$ and $p_k, 0 \leq p_k \leq 1$ is a normalized power coefficient. The superposition of received signals at AP $m$ is

$$r^m = \sqrt{\rho} \sum_{k \in \mathcal{K}} h_k^m \sqrt{p_k} x_k + n^m, \tag{7}$$

where $\rho$ denotes the normalized UL SNR while $n^m \sim \mathcal{CN}(0,1)$ represents the additive noise at AP $m$. Before forwarding the received signals to the CPU for signal detection, the AP applies the matched filtering on its received signals by using the estimate channel. It means that AP $m$ transmits its weighted signal, $\left(\hat{h}_k^m\right)^* r^m$, to the CPU via a perfect backhaul network.[1,2,17] To detect the received signals of user $k$, the CPU uses a receiver filter denoted by $\mathbf{t}_k = \left[t_k^1, t_k^2, \ldots, t_k^M\right]^T$ with $\|t_k\| = 1$ on its received signals.[4] Then, the aggregated received signals at the CPU can be formulated as

$$r_k = \sum_{m\in\mathcal{M}} t_k^m \left(\hat{h}_k^m\right)^* r^m = \sqrt{\rho}\sum_{k'\in\mathcal{K}}\sum_{m\in\mathcal{M}} t_k^m \left(\hat{h}_k^m\right)^* h_{k'}^m \sqrt{p_{k'}} x_{k'} + \sum_{m\in\mathcal{M}} t_k^m \left(\hat{h}_k^m\right)^* n^m. \tag{8}$$

As shown by Ngo et al,[2] as the number of APs is large, the channel hardening effects can be exploited for signal detection. Then, the CPU uses the information of channel statistics rather than CSI. From (8), the aggregated signal at the CPU can be rewritten as

$$r_k = \text{DS}_k x_k + \text{BU}_k x_k + \sum_{k'\in\mathcal{K}\setminus k} \text{IUI}_k^{k'} x_{k'} + \text{TN}_k, \tag{9}$$

where

$$\text{DS}_k = \sqrt{\rho}\,\mathbf{E}\left[\sum_{m\in\mathcal{M}} t_k^m \left(\hat{h}_k^m\right)^* h_k^m \sqrt{p_k}\right], \tag{10a}$$

$$\text{BU}_k = \sqrt{\rho}\left(\sum_{m\in\mathcal{M}} t_k^m \left(\hat{h}_k^m\right)^* h_k^m \sqrt{p_k} - \mathbf{E}\left[\sum_{m\in\mathcal{M}} t_k^m \left(\hat{h}_k^m\right)^* h_k^m \sqrt{p_k}\right]\right) \tag{10b}$$

are the desired signal (DS) and beamforming uncertainty (BU) of user $k$, respectively. Also,

$$\text{IUI}_k^{k'} = \sqrt{\rho}\sum_{m\in\mathcal{M}} t_k^m \left(\hat{h}_k^m\right)^* h_{k'}^m \sqrt{p_{k'}}, \tag{11a}$$

$$\text{TN}_k = \sum_{m\in\mathcal{M}} t_k^m \left(\hat{h}_k^m\right)^* n^m \tag{11b}$$

denote the inter-user interference (IUI) caused by user $k'$ to user $k$ and the total noise (TN), respectively. With the assumption of uncorrelated Gaussian noise,[2] the SINR of user $k$ from (9) can be computed by

$$\text{SINR}_k = \frac{|\text{DS}_k|^2}{\mathbf{E}\left[|\text{BU}_k|^2\right] + \sum_{k'\in\mathcal{K}\setminus k}\mathbf{E}\left[\left|\text{IUI}_k^{k'}\right|^2\right] + \mathbf{E}\left[|\text{TN}_k|^2\right]}. \tag{12}$$

Then, the achievable UL rate of user $k$ is

$$R_k = \log_2(1+\text{SINR}_k) = \log_2\left(1 + \frac{|\text{DS}_k|^2}{\mathbf{E}\left[|\text{BU}_k|^2\right] + \sum_{k'\in\mathcal{K}\setminus k}\mathbf{E}\left[\left|\text{IUI}_k^{k'}\right|^2\right] + \mathbf{E}\left[|\text{TN}_k|^2\right]}\right). \tag{13}$$

Let $\varepsilon_k^m \triangleq h_k^m - \hat{h}_k^m$ be the channel estimation error. Due to the properties of MMSE estimation, $\varepsilon_k^m$ and $\hat{h}_k^m$ are independent. Thus, one has

$$\text{DS}_k = \sqrt{\rho}\sum_{m\in\mathcal{M}}\sqrt{p_k}\,t_k^m\,\mathbf{E}\left[\left(\hat{h}_k^m\right)^* h_k^m\right] = \sqrt{\rho p_k}\sum_{m\in\mathcal{M}} t_k^m \xi_k^m, \tag{14}$$

where $\mathbf{E}\left[\left(\hat{h}_k^m\right)^* \hat{h}_k^m\right] = \mathbf{E}\left[\left|\hat{h}_k^m\right|^2\right] = \xi_k^m$. Hence,

$$|\mathrm{DS}_k|^2 = \rho p_k \left( \sum_{m \in \mathcal{M}} t_k^m \xi_k^m \right)^2. \tag{15}$$

Next, Equation (10b) can be expressed as

$$\mathrm{BU}_k = \sqrt{\rho p_k} \sum_{m \in \mathcal{M}} t_k^m \left( \left(\hat{h}_k^m\right)^* h_k^m - \mathbf{E}\left[\left(\hat{h}_k^m\right)^* h_k^m\right] \right), \tag{16}$$

which can be rewritten as

$$\begin{aligned}\mathbf{E}\left[|\mathrm{BU}_k|^2\right] &= \rho p_k \sum_{m \in \mathcal{M}} (t_k^m)^2 \left( \mathbf{E}\left[\left|\left(\hat{h}_k^m\right)^* h_k^m\right|^2\right] - \left|\mathbf{E}\left[\left(\hat{h}_k^m\right)^* h_k^m\right]\right|^2 \right) \\ &= \rho p_k \sum_{m \in \mathcal{M}} (t_k^m)^2 \left( \mathbf{E}\left[\left|\left(\hat{h}_k^m\right)^* \varepsilon_k^m + \left|\hat{h}_k^m\right|^2\right|^2\right] - (\xi_k^m)^2 \right).\end{aligned} \tag{17}$$

Since $\varepsilon_k^m$ has zero mean and is independent of $\hat{h}_k^m$, $\mathbf{E}\left[\left|\hat{h}_k^m\right|^4\right] = 2(\xi_k^m)^2$, and $\mathbf{E}\left[\left|\varepsilon_k^m\right|^2\right] = \beta_k^m - \xi_k^m$, Equation (17) can be rewritten as[2]

$$\mathbf{E}\{|\mathrm{BU}_k|^2\} = \rho p_k \sum_{m \in \mathcal{M}} (t_k^m)^2 \xi_k^m \beta_k^m. \tag{18}$$

To compute $\mathbf{E}\left[|\mathrm{IUI}_{kk'}|^2\right]$, it is derived from (9) that

$$\mathrm{IUI}_{kk'} = \sqrt{\rho} \sum_{m \in \mathcal{M}} t_k^m \left(\hat{h}_k^m\right) h_{k'}^m \sqrt{p_{k'}}. \tag{19}$$

Substituting (4) into the equation above yields

$$\mathrm{IUI}_{kk'} = \sqrt{\rho} \sum_{m \in \mathcal{M}} t_k^m h_{k'}^m \sqrt{p_{k'}} \vartheta_k^m \left( \sqrt{\tau \rho_\mathrm{p}} \sum_{i \in \mathcal{K}} h_i^m \boldsymbol{\varphi}_k^H \boldsymbol{\varphi}_i + \boldsymbol{\varphi}_k^H [\mathbf{n}_\mathrm{p}]^m \right)^*. \tag{20}$$

Let $\tilde{n}_k^m \triangleq \boldsymbol{\varphi}_k^H [\mathbf{n}_\mathrm{p}]^m \sim \mathcal{CN}(0,1)$. Since $h_i^m$ is independent of $\tilde{n}_k^m$, one has

$$\mathbf{E}\left[|\mathrm{IUI}_{kk'}|^2\right] = \rho p_{k'} \mathbf{E}\left[\left|\sum_{m \in \mathcal{M}} t_k^m h_{k'}^m \vartheta_k^m (\tilde{n}_k^m)^*\right|^2\right] + \rho \tau \rho_\mathrm{p} \mathbf{E}\left\{ p_{k'} \left|\sum_{m \in \mathcal{M}} t_k^m h_{k'}^m \vartheta_k^m \left(\sum_{i \in \mathcal{K}} h_i^m \boldsymbol{\varphi}_k^H \boldsymbol{\varphi}_i\right)^*\right|^2 \right\}. \tag{21}$$

As $\xi_k^m = \mathbf{E}\left[\left|\hat{h}_k^m\right|^2\right] = \sqrt{\tau \rho_\mathrm{p}} \vartheta_k^m \beta_k^m$, the equation above becomes[2]

$$\mathbf{E}\left[|\mathrm{IUI}_{kk'}|^2\right] = \rho p_{k'} \left( \sum_{m \in \mathcal{M}} (t_k^m)^2 \beta_{k'}^m \xi_k^m \right) + \rho p_{k'} |\boldsymbol{\varphi}_k^H \boldsymbol{\varphi}_{k'}|^2 \left( \sum_{m \in \mathcal{M}} t_k^m \xi_k^m \frac{\beta_{k'}^m}{\beta_k^m} \right)^2. \tag{22}$$

Now, compute $\mathbf{E}\{|\mathrm{TN}_k|^2\}$. From (9), as $\hat{h}_k^m$ and $n_m$ are uncorrelated, the TN for user $k$ is given by

$$\mathbf{E}\{|\mathrm{TN}_k|^2\} = \sum_{m \in \mathcal{M}} (t_k^m)^2 \xi_k^m. \tag{23}$$

By substituting (15), (18), (22), and (23) into (12), the closed form of the SINR of user $k$ is given as follows:

$$\text{SINR}_k = \frac{\rho p_k \left(\sum_{m \in \mathcal{M}} t_k^m \xi_k^m\right)^2}{\sum_{k' \in \mathcal{K} \setminus k} \rho p_{k'} |\boldsymbol{\varphi}_k^H \boldsymbol{\varphi}_{k'}|^2 \left(\sum_{m \in \mathcal{M}} t_k^m \xi_k^m \frac{\beta_{k'}^m}{\beta_k^m}\right)^2 + \sum_{k' \in \mathcal{K}} \rho p_{k'} \left(\sum_{m \in \mathcal{M}} (t_k^m)^2 \beta_{k'}^m \xi_k^m\right) + \sum_{m \in \mathcal{M}} (t_k^m)^2 \xi_k^m}. \quad (24)$$

We define $\boldsymbol{\xi}_k \triangleq [\xi_k^1, \xi_k^2, ..., \xi_k^M]^T$, $\mathbf{t}_k \triangleq [t_k^1, t_k^2, ..., t_k^M]^T$, $\boldsymbol{\beta}_k \triangleq [\beta_k^1, \beta_k^2, ..., \beta_k^M]^T$. Also, let $\boldsymbol{\varsigma}_k^{k'} = |\boldsymbol{\varphi}_k^H \boldsymbol{\varphi}_{k'}|^2 \boldsymbol{\xi}_k \odot \boldsymbol{\beta}_{k'} \oslash \boldsymbol{\beta}_k$, $\boldsymbol{\Upsilon}_k^{k'} = \text{diag}(\boldsymbol{\xi}_k \odot \boldsymbol{\beta}_{k'})$, $\boldsymbol{\Xi}_k = \text{diag}(\boldsymbol{\xi}_k)$. Then, the achievable UL rate of user $k$ can be expressed as

$$R_k = \log_2 \left(1 + \frac{\mathbf{t}_k^H (\rho p_k \boldsymbol{\xi}_k \boldsymbol{\xi}_k^H) \mathbf{t}_k}{\mathbf{t}_k^H \left(\sum_{k' \in \mathcal{K} \setminus k} \rho p_{k'} \boldsymbol{\varsigma}_k^{k'} (\boldsymbol{\varsigma}_k^{k'})^H + \sum_{k' \in \mathcal{K}} \rho p_{k'} \boldsymbol{\Upsilon}_k^{k'} + \boldsymbol{\Xi}_k\right) \mathbf{t}_k}\right). \quad (25)$$

To improve trade-off between the achievable sum rate and the fairness among users, in this paper, we aim at maximizing the PF utility function which is defined as the sum of the logarithm of the user achievable rate.[9,10] By denoting $\mathbf{t} \triangleq (\mathbf{t}_k)_{k=1,...,K}, p \triangleq (p_k)_{k=1,...,K}$, the design problem of the receiver filter coefficients at the CPU and power allocation at users can be mathematically formulated as

$$\mathcal{P}_1: \max_{\mathbf{t},p} f(\mathbf{t},p) \triangleq \sum_{k \in \mathcal{K}} \log_2(R_k) \quad (26a)$$

$$||\mathbf{t}_k|| = 1, k = 1,...,K, \quad (26b)$$

$$0 \le p_k \le 1. \quad (26c)$$

It can be seen that problem $\mathcal{P}_1$ is nonlinear and nonconvex optimization since the coupling between the receiver filter coefficients (i.e., $t$) and transmit power variables (i.e., $p$). In the next section, we will introduce an iterative algorithm to solve problem (26).

## 3 | PROPOSED ITERATIVE ALGORITHM FOR PF

To handle with the coupling of the design variables in problem $\mathcal{P}_1$, we adopt an alternative optimization approach in which either the filter coefficients or the power allocations are found while the other are fixed.

### 3.1 | Receiver filter coefficient design

It is observed from (25) that the achievable rate of user $k$ only relies on the filter coefficient $t_k$ (does not reply on the other filter coefficients). Thus, the optimal filter coefficients in problem (26) can be obtained by maximizing the SINRs of the users individually.[4,18-20] The optimal coefficients $t_k$ of user $k$ can be achieved by solving the following maximization problem:

$$\mathcal{P}_2: \max_{\mathbf{t}_k} \text{SINR}_k \quad (27a)$$

$$s.t. ||\mathbf{t}_k|| = 1 \quad (27b)$$

From (25), we have

$$\text{SINR}_k = \frac{\mathbf{t}_k^H \left( \rho p_k \boldsymbol{\xi}_k \boldsymbol{\xi}_k^H \right) \mathbf{t}_k}{\mathbf{t}_k^H \left( \sum_{k' \in \mathcal{K} \backslash k} \rho p_{k'} \boldsymbol{\varsigma}_k^{k'} \left( \boldsymbol{\varsigma}_k^{k'} \right)^H + \sum_{k' \in \mathcal{K}} \rho p_{k'} \Upsilon_k^{k'} + \boldsymbol{\Xi}_k \right) \mathbf{t}_k}. \tag{28}$$

By letting

$$\mathbf{N}_k = \rho p_k \boldsymbol{\xi}_k \boldsymbol{\xi}_k^H, \tag{29a}$$

$$\mathbf{D}_k = \sum_{k' \in \mathcal{K} \backslash k} \rho p_{k'} \boldsymbol{\varsigma}_k^{k'} \left( \boldsymbol{\varsigma}_k^{k'} \right)^H + \sum_{k' \in \mathcal{K}} \rho p_{k'} \Upsilon_k^{k'} + \boldsymbol{\Xi}_k, \tag{29b}$$

problem $\mathcal{P}_2$ can be rewritten as

$$\max_{\mathbf{t}_k} \quad \frac{\mathbf{t}_k^H \mathbf{N}_k \mathbf{t}_k}{\mathbf{t}_k^H \mathbf{D}_k \mathbf{t}_k} \tag{30a}$$

$$s.t. ||\mathbf{t}_k|| = 1. \tag{30b}$$

By defining

$$\mathbf{t}_k = \frac{\hat{\mathbf{t}}_k}{||\hat{\mathbf{t}}_k||} \tag{31}$$

problem (30) can be equivalently rewritten as

$$\max_{\hat{\mathbf{t}}_k} \quad \hat{\mathbf{t}}_k^H \mathbf{N}_k \hat{\mathbf{t}}_k \tag{32a}$$

$$\text{s.t.} \quad \hat{\mathbf{t}}_k^H \mathbf{D}_k \hat{\mathbf{t}}_k = 1. \tag{32b}$$

The Lagrangian[21] for (32) is

$$\mathcal{L}(\hat{\mathbf{t}}_k, \lambda) = \hat{\mathbf{t}}_k^H \mathbf{N}_k \hat{\mathbf{t}}_k - \lambda \left( \hat{\mathbf{t}}_k^H \mathbf{D}_k \hat{\mathbf{t}}_k - 1 \right), \tag{33}$$

where $\lambda \in \mathbb{R}$ is the Lagrangian multiplier. By equating the derivative of Lagrangian to zero, we have

$$\frac{\partial \mathcal{L}(\hat{\mathbf{t}}_k, \lambda)}{\partial \hat{\mathbf{t}}_k} = 2\mathbf{N}_k \hat{\mathbf{t}}_k - 2\lambda \mathbf{D}_k \hat{\mathbf{t}}_k = 0, \Leftrightarrow \mathbf{N}_k \hat{\mathbf{t}}_k = \lambda \mathbf{D}_k \hat{\mathbf{t}}_k. \tag{34}$$

Therefore, considering the pair of matrix $\mathbf{N}_k$ and $\mathbf{D}_k$, $\lambda$ and $\hat{\mathbf{t}}_k$ are a generalized eigenvalue and a generalized eigenvectors, respectively. This means that $\lambda \in \left\{ \tilde{\lambda} \in \mathbb{R} : \det \left( \mathbf{N}_k - \tilde{\lambda} \mathbf{D}_k \right) = 0 \right\}$. Since problem (32) is a maximization problem, the optimal vector $\hat{\mathbf{t}}_k^*$ is the eigenvector which corresponds to the maximal eigenvalue. Then, the optimal filter coefficients $\mathbf{t}_k^*$ to problem (30) are

$$\mathbf{t}_k^* = \frac{\hat{\mathbf{t}}_k^*}{||\hat{\mathbf{t}}_k^*||} \tag{35}$$

The solution to problem $\mathcal{P}_2$ is summarized in Algorithm 1.

**Algorithm 1** Receiver filter coefficient design for user $k$.

**Input**: $N_k; D_k$
1. Compute the largest eigenvalue from $\det(N_k - \lambda D_k) = 0$ and a corresponding eigenvector $\hat{t}_k^*$.
2. Compute
$$t_k^* = \frac{\hat{t}_k^*}{\|\hat{t}_k^*\|}$$

**Output**: $t_k^*$

## 3.2 | Power allocation

Next, we address the problem of allocating transmit powers to the users by fixing the receiver filter coefficients $t$ which are achieved through Algorithm 1. Then, the optimization problem (26) can be written as

$$\mathcal{P}_3: \max_p \; f_{\mathbf{t}^*}(p) \triangleq \sum_{k \in \mathcal{K}} \log_2(R_k) \tag{36a}$$

$$\text{s.t.} \; 0 \leq p_k \leq 1, \; \forall k. \tag{36b}$$

To deal with the challenges associated with the nonlinear and nonconvex optimization problem (36), we develop a GP algorithm. In addition to two essential stages of the GP algorithm, namely, computing gradients and projecting the updated variables, we adopt Armijo's rule[13,14] to guarantee the convergence of the iterative algorithm. The algorithm to solve problem $\mathcal{P}_3$ is represented as follows.

### 3.2.1 | Computing the gradients

Given the filter coefficients, the SINR can be compactly rewritten as

$$\text{SINR}_k = \frac{\alpha_k p_k}{\sum_{k' \in \mathcal{K} \setminus k} \eta_k^{k'} p_{k'} + \sum_{k' \in \mathcal{K}} \chi_k^{k'} p_{k'} + \delta_k}, \tag{37}$$

where $\alpha_k = \rho \mathbf{t}_k^H \xi_k \xi_k^H \mathbf{t}_k$, $\eta_k^{k'} = \rho \mathbf{t}_k^H \varsigma_k^{k'} (\varsigma_k^{k'})^H \mathbf{t}_k$, $\chi_k^{k'} = \rho \mathbf{t}_k^H \Upsilon_k^{k'} \mathbf{t}_k$, $\delta_k = \mathbf{t}_k^H \Xi_k \mathbf{t}_k$. In addition, by defining

$$g_k(p) = \log_2(\log_2(1 + \text{SINR}_k)) = \log_2(e) \ln \left( \underbrace{\ln(1 + \text{SINR}_k)}_{\hat{R}_k} \right) + \log_2(e) \ln(\log_2(e)), \tag{38}$$

the objective function in (36a) can be rewritten as follows:

$$f_{\mathbf{t}^*}(p) = \sum_{k \in \mathcal{K}} g_k(p). \tag{39}$$

Therefore, the gradient of the objective function with respect to $p_i, i \in \mathcal{K}$ is then

$$\nabla_{p_i} f_{\mathbf{t}^*}(p) \triangleq \frac{\partial f_{\mathbf{t}^*}(p)}{\partial p_i} = \frac{\partial g_i(p)}{\partial p_i} + \sum_{k \in \mathcal{K} \setminus i} \frac{\partial g_k(p)}{\partial p_i}. \tag{40}$$

From Equation (38), we have

$$\frac{\partial g_k(p)}{\partial p_i} = \log_2(e) \frac{1}{\hat{R}_k} \frac{\partial \hat{R}_k}{\partial p_i}. \tag{41}$$

With $k = i$, we can derive

$$\frac{\partial \hat{R}_i}{\partial p_i} = \frac{\partial}{\partial p_i} \ln\left(1 + \frac{\alpha_i p_i}{\sum_{k' \in \mathcal{K} \setminus i} \eta_i^{k'} p_{k'} + \sum_{k' \in \mathcal{K}} \chi_i^{k'} p_{k'} + \delta_i}\right) = \frac{\alpha_i + \chi_i^i}{\alpha_i p_i + \sum_{k' \in \mathcal{K} \setminus i} \eta_i^{k'} p_{k'} + \sum_{k' \in \mathcal{K}} \chi_i^{k'} p_{k'} + \delta_i} - \frac{\chi_i^i}{\sum_{k' \in \mathcal{K} \setminus i} \eta_i^{k'} p_{k'} + \sum_{k' \in \mathcal{K}} \chi_i^{k'} p_{k'} + \delta_i}. \tag{42}$$

With $k \neq i$, we have

$$\frac{\partial \hat{R}_k}{\partial p_i} = \frac{\partial}{\partial p_i} \ln\left(1 + \frac{\alpha_k p_k}{\sum_{k' \in \mathcal{K} \setminus k} \eta_k^{k'} p_{k'} + \sum_{k' \in \mathcal{K}} \chi_k^{k'} p_{k'} + \delta_k}\right) = \frac{\eta_k^i + \chi_k^i}{\alpha_k p_k + \sum_{k' \in \mathcal{K} \setminus k} \eta_k^{k'} p_{k'} + \sum_{k' \in \mathcal{K}} \chi_k^{k'} p_{k'} + \delta_k} - \frac{\eta_k^i + \chi_k^i}{\sum_{k' \in \mathcal{K} \setminus k} \eta_k^{k'} p_{k'} + \sum_{k' \in \mathcal{K}} \chi_k^{k'} p_{k'} + \delta_k}. \tag{43}$$

From (41) and (42), we have

$$\frac{\partial g_i(p)}{\partial p_i} = \log_2(e) \frac{1}{\hat{R}_i} \left( \frac{\alpha_i + \chi_i^i}{\alpha_i p_i + \sum_{k' \in \mathcal{K} \setminus i} \eta_i^{k'} p_{k'} + \sum_{k' \in \mathcal{K}} \chi_i^{k'} p_{k'} + \delta_i} - \frac{\chi_i^i}{\sum_{k' \in \mathcal{K} \setminus i} \eta_i^{k'} p_{k'} + \sum_{k' \in \mathcal{K}} \chi_i^{k'} p_{k'} + \delta_i} \right). \tag{44}$$

From (41) and (43), with $k \neq i$, we have

$$\frac{\partial g_k(p)}{\partial p_i} = \log_2(e) \frac{1}{\hat{R}_k} \left( \frac{\eta_k^i + \chi_k^i}{\alpha_k p_k + \sum_{k' \in \mathcal{K} \setminus k} \eta_k^{k'} p_{k'} + \sum_{k' \in \mathcal{K}} \chi_k^{k'} p_{k'} + \delta_k} - \frac{\eta_k^i + \chi_k^i}{\sum_{k' \in \mathcal{K} \setminus k} \eta_k^{k'} p_{k'} + \sum_{k' \in \mathcal{K}} \chi_k^{k'} p_{k'} + \delta_k} \right). \tag{45}$$

By substituting (44) and (45) into (40), we obtain the gradients of the objective function. The power coefficients are then updated by

$$p_i' = p_i + \delta \nabla_{p_i} f_{\mathbf{t}^*}(p), \tag{46}$$

where $\delta$ is the step size.

### 3.2.2 | Finding the projections

To guarantee the updated variables satisfying the power constraint (36b), we project $p_i'$ into the feasible region. Let $\bar{p} = \mathcal{P}(p', 0, 1)$ be the projection of $p'$ onto $[0; 1]$. The projection process can be rewritten as

$$\bar{p}_i = \mathcal{P}(p_i', 0, 1) = \begin{pmatrix} 0 & \text{if } p_i' < 0 \\ p_i' & \text{if } 0 \leq p_i' \leq 1 \\ 1 & \text{if } 1 \leq p_i' \end{pmatrix}. \tag{47}$$

### 3.2.3 | Applying Armijo's rule

After projecting the power coefficients, we use Armijo's rule with provable convergence to update the coefficients for the next iteration.[14,22] With fixed values $0 \leq \sigma \leq 1$ and $0 \leq \beta \leq 1$, the updated coefficients for the $(n+1)$ th iteration are chosen as

$$p_k^{(n+1)} = p_k^{(n)} + \beta^{m_n}\left(\bar{p}_k^{(n)} - p_k^{(n)}\right), \tag{48}$$

where $m_n$ is the first nonnegative integer that satisfies

$$f_{\mathbf{t}^*}\left(p^{(n+1)}\right) - f_{\mathbf{t}^*}\left(p^{(n)}\right) \geq \sigma\beta^{m_n}\sum_{k\in\mathcal{K}}\nabla_{p_k}f_{\mathbf{t}^*}\left(p^{(n)}\right)\left(\bar{p}_k^{(n)} - p_k^{(n)}\right). \tag{49}$$

The iterative algorithm to obtain the optimal filter coefficients is described in Algorithm 2.

---

**Algorithm 2** :GP algorithm for power allocation in the $\kappa$th iteration.

**Initialize**:
- Input $p^{(\kappa)}, t^{(\kappa+1)}$.
- $n = 1$.
- *search* = true

**repeat**
    1. Calculate the gradients $\quad\nabla_{p_k}f_{\mathbf{t}^*}\left(q^{(n)}\right)$.

    2. Compute $\quad p'^{(n)}_k = p^{(n)}_k + \delta\nabla_k f_{\mathbf{t}^*}\left(p^{(n)}\right)$.

    3. Find the projections $\quad\bar{p}^{(n)}_k$ of $p'^{(n)}_k$ onto $[0;1]$.

    4. Applying the Armijo's rule to find $m_n$.
$m_n = 0$
    **repeat**
        Computing:
$p_k^{(n+1)} = p_k^{(n)} + \beta^{m_n}\left(\bar{p}_k^{(n)} - p_k^{(n)}\right)$
$m_n = m_n + 1$
    **until** Satisfying (49)
    5. Considering terminating condition.
    **if** $\frac{f(t^{(\kappa+1)},p^{(n+1)}) - f(t^{(\kappa+1)},p^{(n)})}{f(t^{(\kappa+1)},p^{(n)})} \leq \varepsilon$ **then**
        *search* = false
    **else**
        $n = n + 1$
    **end if**
**until** *search* = false
**Output**: $p^{(\kappa+1)} = p^{(n+1)}$.

---

As a result, the iterative optimization algorithm to obtain the solution to problem $\mathcal{P}_1$ is to alternatively solve Algorithms 1 and 2. These subproblems will be solved iteratively until

$$\frac{f\left(\mathbf{t}^{(\kappa+1)},p^{(\kappa+1)}\right) - f\left(\mathbf{t}^{(\kappa)},p^{(\kappa)}\right)}{f\left(\mathbf{t}^{(\kappa)},p^{(\kappa)}\right)} \leq \varepsilon, \tag{50}$$

where $\varepsilon$ is the tolerance error. The algorithm solving problem $\mathcal{P}_1$ in (26) is represented in Algorithm 3.

**Algorithm 3**: Iterative algorithm for the SE PF maximization problem $\mathcal{P}_1$ in (26).

**Initialize**:
- Set $\kappa = 1$, $search = \text{true}$, $\varepsilon$.
- Choose initial variables: $\boldsymbol{\varphi}_k, p_k^{(1)}, \quad \forall k$.

**repeat**

1. Solve $\mathcal{P}_2$ in (27) with fixed $p_k^{(\kappa)}$ by using Algorithm 1 to obtain $\boldsymbol{t}^{*(\kappa)}$.

2. Solve $\mathcal{P}_3$ in (36) with fixed $\boldsymbol{t}_k^{*(\kappa)}$ by using Algorithm 2 to obtain $p^{(\kappa+1)}$.

3. Considering terminating condition in (50).
    **if** $\frac{f(\boldsymbol{t}^{(\kappa+1)}, p^{(\kappa+1)}) - f(\boldsymbol{t}^{(\kappa)}, p^{(\kappa)})}{f(\boldsymbol{t}^{(\kappa)}, p^{(\kappa)})} \leq \varepsilon$ **then**
        $search = \text{false}$
    **else**
        $\kappa = \kappa + 1$
    **end if**
**until** $search = \text{false}$
**Output**: $\boldsymbol{t}^{(\kappa+1)}, p^{(\kappa+1)}$.

## 3.3 | Convergence analysis

In this subsection, we will prove the convergence of Algorithm 3, which leads to a locally optimal solution to problem $\mathcal{P}_1$. First, we prove the convergence of the proposed GP method in Algorithm 2. Given feasible solutions $0 \leq p_k^{(n)} \leq 1$, Equation (46) at the $(n+1)$ iteration can be written as

$$p_k^{\prime(n)} = p_k^{(n)} + \delta \nabla_k f_{\boldsymbol{t}^*}\left(p^{(n)}\right). \quad (51)$$

We consider all possible following cases:

- If $0 \leq p_k^{\prime(n)} \leq 1$, the projection in Equation (47) gives $\bar{p}_k^{(n)} = p_k^{\prime(n)}$, which results in $\bar{p}_k^{(n)} - p_k^{(n)} = \delta \nabla_k f_{\boldsymbol{t}^*}\left(p^{(n)}\right)$. Thus, we have $\nabla_{p_k} f_{\boldsymbol{t}^*}(p^{(n)})\left(\bar{p}_k^{(n)} - p_k^{(n)}\right) \geq 0$.
- If $p_k^{\prime(n)} \leq 0$, Equation (51) implies that $\nabla_k f_{\boldsymbol{t}^*}(p^{(n)}) \leq 0$ while the projection in Equation (47) yields $\bar{p}_k^{(n)} = 0$. This leads to $\nabla_{p_k} f_{\boldsymbol{t}^*}(p^{(n)})\left(\bar{p}_k^{(n)} - p_k^{(n)}\right) \geq 0$.
- If $p_k^{\prime(n)} \geq 1$, Equation (51) implies that $\nabla_k f_{\boldsymbol{t}^*}(p^{(n)}) \geq 0$ while the projection Equation (47) yields $\bar{p}_k^{(n)} = 1$. Thus, we have $\nabla_{p_k} f_{\boldsymbol{t}^*}(p^{(n)})\left(\bar{p}_k^{(n)} - p_k^{(n)}\right) \geq 0$.

In other words, the right-hand side of Equation (49) is always nonnegative, and thus, the objective function in (36a) is nondecreasing over iterations. In addition, with the constrain (36b), the objective function (36a) is upper bounded. Consequently, by applying iterative Algorithm 2, the objective function in problem $\mathcal{P}_3$ will converge to at least a locally optimal solution.

Next, based on the convergence of Algorithm 2, we will prove the convergence of Algorithm 3. Let $f\left(\boldsymbol{t}^{(\kappa)}, p^{(\kappa)}\right)$ be the value of the objective function in problem $\mathcal{P}_1$ obtained at the $\kappa$th iteration of Algorithm 3. In the $(\kappa + 1)$ iteration of Algorithm 3, we have

$$f\left(\boldsymbol{t}^{(\kappa)}, p^{(\kappa)}\right) \leq f\left(\boldsymbol{t}^{(\kappa+1)}, p^{(\kappa)}\right), \quad (52)$$

where $\boldsymbol{t}^{(\kappa+1)}$ are updated receiver filter coefficients obtained by solving problem $\mathcal{P}_2$ with fixed $p^{(\kappa)}$. Also, by the nondecreasing of objective function in Algorithm 2, one has

$$f\left(\mathbf{t}^{(\kappa+1)}, p^{(\kappa)}\right) \leq f\left(\mathbf{t}^{(\kappa+1)}, p^{(\kappa+1)}\right), \quad (53)$$

where $p^{(\kappa+1)}$ are local optimal transmit powers achieved by solving Problem $\mathcal{P}_3$ with fixed $\mathbf{t}^{(\kappa+1)}$. Therefore, from Equation (52) and (53), after the $(\kappa+1)$th iteration, we have

$$f\left(\mathbf{t}^{(\kappa)}, p^{(\kappa)}\right) \leq f\left(\mathbf{t}^{(\kappa+1)}, p^{(\kappa+1)}\right), \quad (54)$$

This means that the objective function in problem $\mathcal{P}_1$ is nondecreasing through each iteration of Algorithm 3. Also, as the receiver filter coefficients and transmit powers are imposed by constraints (26b) and (26c), the objective function is upper bounded and will converge to a local optimal value over iterations. In addition, the convergence of Algorithm 3 can be verified through simulation in Section 4.

## 3.4 | Complexity analysis

The computational complexity of Algorithm 3 comes from that of Algorithms 1 and 2. In each iteration of Algorithm 1, the computational complexity of solving the eigenvalue problem $\mathcal{P}_2$ for $K$ users is $\mathcal{O}(KM^3)$. Regarding Algorithm 2, the computational complexity of computing the gradient in Equation (40) is $\mathcal{O}(KM^2 + K^2M)$. Also, the computational complexity of applying Armijo's rule is $\mathcal{O}(mK)$,[14] where $m$ is the number of trials in Armijo's rule. Assuming that Algorithm 2 takes $n_{\max}$ iterations to converge, the complexity of the algorithm is $\mathcal{O}(n_{\max}(KM^2 + K^2M + mK))$. Therefore, the computational complexity of Algorithm 3 is $\mathcal{O}(\kappa_{\max}(KM^3 + n_{\max}(KM^2 + K^2M + mK)))$ where $\kappa_{\max}$ is the maximum iteration number of Algorithm 3.

## 4 | NUMERICAL RESULTS

In this section, numerical simulation results are provided to examine the performance of the proposed algorithm for the CF massive MIMO network. We consider the CF massive MIMO network with $M$ single-antenna APs and $K$ single-antenna users, which are randomly distributed in a $D \times D$ square area.[2] The noise power $\rho_n$ is modeled as

$$\rho_n = B \times k_B \times T_0 \times W, \quad (55)$$

where B=20 MHz is the bandwidth, $k_B$=1.381×10$^{-23}$ J.K$^{-1}$ denotes the Boltzmann constant, $T_0$=290° K represents the noise temperature, and $W$=9 dB stands for the noise figure. The transmit powers of pilot sequences and data symbols are, respectively, denoted as $\bar{\rho}_p$ and $\bar{\rho}$. Then, we have

$$\rho_p = \frac{\bar{\rho}_p}{\rho_n}, \quad \rho = \frac{\bar{\rho}}{\rho_n}. \quad (56)$$

**TABLE 1** System parameters

| Parameter | Value |
| --- | --- |
| Carrier frequency | 1.9 GHz |
| Bandwidth | 20 MHz |
| AP antenna height | 15 m |
| User antenna height | 1.65 m |
| Noise figure | 9 dB |
| $\bar{\rho}_p, \bar{\rho}$ | 200 mW |
| $\sigma_{sh}$ | 8 dB |
| $D, d_1, d_0$ | 1000, 50, and 10 m |

We set $\bar{\rho}_p = \bar{\rho} = 200$ mW. Large-scale fading $\beta_k^m$ between user $k$ and AP $m$ in (1) is modeled as[2]

$$\beta_k^m = 10^{\frac{\text{PL}_k^m}{10}} 10^{\frac{\sigma_{sh} z_k^m}{10}}, \tag{57}$$

where $\text{PL}_k^m$ is the path loss, $10^{\frac{\sigma_{sh} z_k^m}{10}}$ represents shadow fading with standard deviation $\sigma_{sh} = 8$ dB, and $z_k^m \sim \mathcal{N}(0,1)$. In the simulation, we consider an uncorrelated shadowing model with three slopes of the path loss which are given as[2,23]

$$\text{PL}_k^m = \begin{pmatrix} -L - 35\log_{10}(d_k^m), & \text{if } d_1 < d_k^m \\ -L - 15\log_{10}(d_1) - 20\log_{10}(d_k^m), & \text{if } d_0 < d_k^m \leq d_1 \\ -L - 15\log_{10}(d_1) - 20\log_{10}(d_0), & \text{if } d_k^m \leq d_0 \end{pmatrix}, \tag{58}$$

where $d_k^m$ is the distance from user $k$ to AP $m$ (in km). Here, $L$ is computed by

$$L \triangleq 46.3 + 33.9\log_{10}(f) - 13.82\log_{10}(h_{\text{AP}}) - (1.1\log_{10}(f) - 0.7)h_u + (1.56\log_{10}(f) - 0.8), \tag{59}$$

where $f$ is the carrier frequency (in MHz), $h_{\text{AP}}$ is the AP antenna height (in m), and $h_u$ denotes the user antenna height (in m). The simulation area is wrapped around at the edges to reproduce an infinity coverage area.[2] The simulation parameters are summarized in Table 1. To take the overheads of the UL channel estimation and the DL data transmission into account, the net throughput is defined as

$$R_k^{\text{net}} = \frac{1 - \tau/\tau_c}{2} R_k. \tag{60}$$

All simulation results are averaged over 200 random realizations of the locations of APs, users, and shadow fading. Similar to the work,[22] the parameters for iterative algorithms are set to $\delta = 1, \sigma = 0.1$, and $\beta = 0.5$. For the tolerance error $\varepsilon$, our attempts for $\varepsilon = 10^{-3}$ and $\varepsilon = 10^{-5}$ yield almost the same average achievable rates. Thus, for the sake of efficiency of computational complexity, the tolerance error is set to $\varepsilon = 10^{-3}$ in our simulations.

**Example 1** In this example, we investigate the convergence characteristics of Algorithm 3 for a CF massive MIMO network consisting of $M$=120 APs, $K$=20 users, $\tau$=10, and $\tau_c$=200 with various realizations. Figure 2 presents values of the objective function in $\mathcal{P}_1$ over iterations for four random realizations. As can be seen from the figure, the sum of logarithm net rates increases through iterations and converges after only a few iterations.

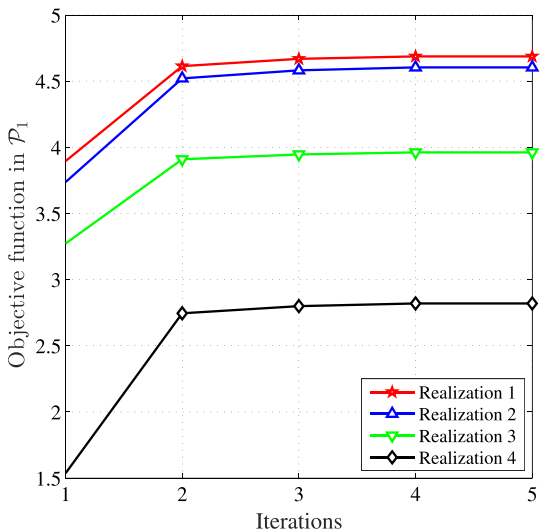

**FIGURE 2** The convergence of the objective function in $\mathcal{P}_1$ versus iteration $\kappa$ in Algorithm 3 with $M = 120, K = 20$, and $\tau = 10$

**Example 2** This example examines the average achievable sum rates for the CF massive MIMO networks with two scenarios with $K = 40$ users and $K = 20$ users for various number of APs. For comparison purposes, we also include the average achievable sum rates of the max-min SINR approach proposed by Bashar et al[4] and SRM. We also use the proposed GP method for the SRM. Figure 3 presents the achievable sum rates versus the number of APs with $\tau = 10$, $\tau_c = 200$. It can be seen from Figure 3 that the sum rate achieved from solving the SEPFM and the SRM problems are nearly the same. Also, it can be observed that the sum rate of a network can be improved by increasing either the number of APs or the number of users in the network. The achievable sum rates will be significantly improved when the number of APs is much more greater than that of users. For example, with a modest number of APs, the sum-rate differences between the 40-user and the 20-user networks are less recognizable. In contrast, at noticeable number-of-antenna regimes, the differences are much more recognizable.

**Example 3** In this example, we study the fairness of the achievable rates among users. Figure 4 shows the average maximum user rates and average minimum user rates in the networks with $K = 20$ and $\tau = 10$ with three approaches: SEPFM, SRM, and max-min SINR.[4] It can be seen from Figure 4 that each approach outcomes a specific gap between the maximum and the minimum rates in a network. In particular, while the SRM approach results in the largest gaps between the maximal rates and the minimal user rates, the max-min SINR approach[4] results in an equal rate for all users in the network, which leads to an overlap of the maximum and the minimum user rate curves. To the SEPFM, its curves of maximal and minimal user rates are in the middle of those of the SRM. It means that the SEPFM provides good balance between the sum rate and user rate fairness. Also, from Figure 4, the achievable rates of either the maximum or the minimum user rates tend to increase with the rising number of the APs.

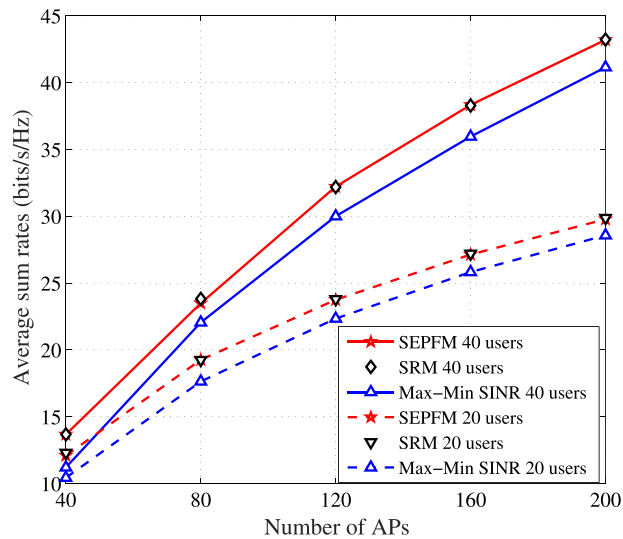

**FIGURE 3** Achievable sum rate with $\tau = 10$

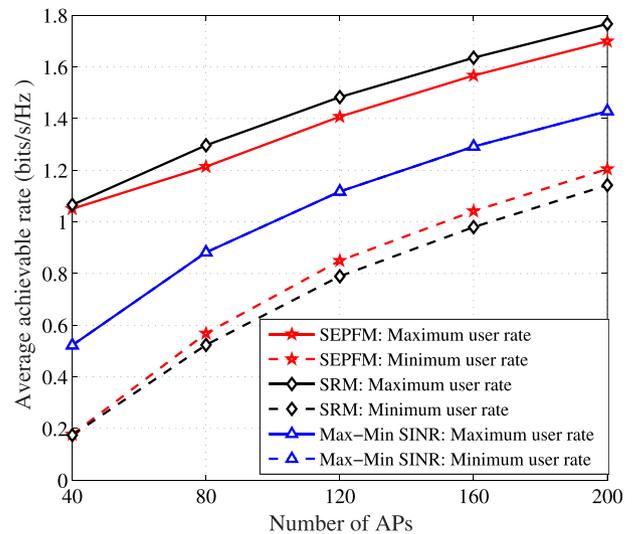

**FIGURE 4** Average maximum and minimum user rates with $K = 20$ and $\tau = 10$

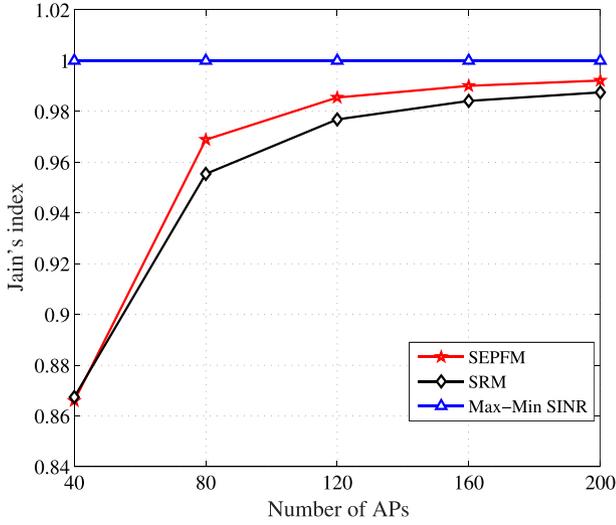

**FIGURE 5** Jain's index with $K=20$ and $\tau=10$

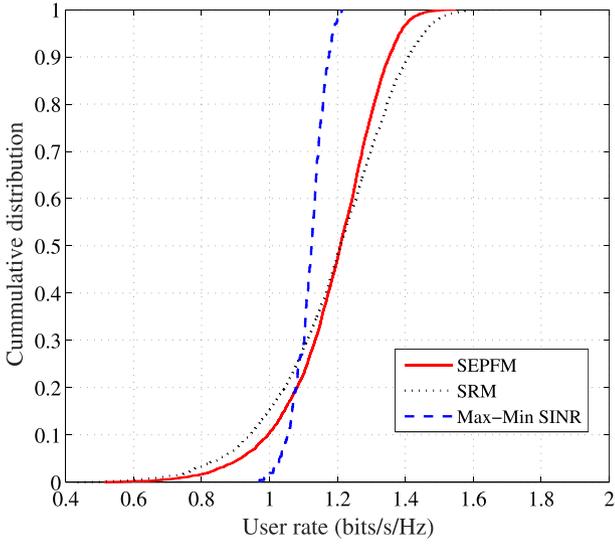

**FIGURE 6** The cumulative distribution of the users' rates with $M=120, K=20$, and $\tau=10$

To numerically analyze the fairness, we adopt Jain's index[24] given by

$$J = \frac{\left(\sum_{k\in\mathcal{K}} R_k^{\text{net}}\right)^2}{K\sum_{k\in\mathcal{K}} \left(R_k^{\text{net}}\right)^2}. \quad (61)$$

It has been suggested that Jain's index in (61) ranges from $\frac{1}{K}$ to 1. Particularly, when $J = \frac{1}{K}$, the network has the least fair rates among users, which means that the transmission is performed by only user with a good channel condition. In contrast, when $J = 1$, the network provides the fairest rates to its users. In this case, all users are supposed to have the same rates, which can be achieved by the max-min SINR approach.[2,4] Figure 5 shows the Jain's indices of the networks with $K = 20$ and $\tau = 10$ with three different approaches of the SEPFM, SRM, and max-min SINR.[4] Obviously, the max-min SINR[4] power control provides the most fairness as this approach results in the same rates for all users in a network. While the SEPFM scheme leads to reasonable fairness, the SRM method provides the lowest indices. Also, the indices tend to converge to 1 when increasing the number of APs. It can be seen from Figure 6 that, at 80 APs, both the SEPFM and the SRM approaches provide the indices that are greater than 12% below 1. However, the indices are approximately only 1% less than 1 at 200 APs.

To obtain the insights into the achievable rates, Figure 6 presents the cumulative distribution of the users' rates for three approaches[4] with $M = 120, K = 20$, and $\tau = 10$. It can be observed from Figure 6 that as compared with the SRM method, the achievable user rate of our SEPFM method is more concentrated around its median. However, it is less concentrated around the median in comparison with the max-min SINR. On the other hand, the median of the achievable rate provided by the proposed algorithm is nearly identical to that of the SRM method while it is higher than the median achieved by the max-min SINR approach.[4] More specifically, at least 50% of the achievable rates provided by the proposed algorithm are higher than those achieved by the max-min SINR approach. These results confirmed that our SEPFM can provide a good balance between the achievable sum rate and the user rate fairness.

## 5 | CONCLUSIONS

We have considered a CF massive MIMO which is a potential technology for future wireless networks. We have studied the PF of UL SE maximization problem in CF massive MIMO systems. To handle the challenges in solving the non-convexity optimization problem that jointly designs the receiver filter coefficients at the CP and power allocation at users, we have proposed an iterative algorithm. The primary maximization problem is decomposed into two subproblems which is addressed by solving a generalized eigenvalue problem and deploying a GP algorithm. Simulation results have been provided to illustrate the effectiveness of our proposed algorithm with various circumstances. By embedding the PF notion into the utility function of the SE, the proposed algorithm offers a good balance between maximizing the sum of user rates and maintaining user rate fairness among them.